\documentclass{bioinfo}
\copyrightyear{2021} \pubyear{2021}

\access{Advance Access Publication Date: Day Month Year}
\appnotes{Manuscript Category}

\usepackage{algorithm}
\usepackage{graphicx}
\usepackage{amsmath}
\usepackage{subfigure}
\usepackage{soul,color}
\usepackage{algorithm,algpseudocode}
\usepackage{url}
\usepackage{breqn}
\usepackage{caption}
\usepackage{amsthm}
\theoremstyle{definition}

\newtheorem{definition}{Definition}[section]
\usepackage{titlesec}
\usepackage{xcolor}

\begin{document}

\bibliographystyle{bioinformatics}

\firstpage{1}
\subtitle{Genome Analysis}

\title[Near-Optimal Privacy-Utility Tradeoff in Genomic Studies Using Selective SNP Hiding]{Near-Optimal Privacy-Utility Tradeoff in Genomic Studies Using Selective SNP Hiding}
\author[Almadhoun Alserr \textit{et~al}.]{Nour Almadhoun Alserr\,$^{\text{\sfb 1}}$, Gulce Kale\,$^{\text{\sfb 3}}$, Onur Mutlu\,$^{\text{\sfb 1}*}$, Oznur Tastan\,$^{\text{\sfb 4}*}$, and Erman Ayday\,$^{\text{\sfb 2,3,}*}$}
\address{$^{\text{\sf 1}}$Department of Information Technology and Electrical Engineering, ETH Zurich, Zurich 8006, Switzerland\,  $^{\text{\sf 2}}$Department of Electrical Engineering and Computer Science, Case Western Reserve University, Cleveland, OH 44106, USA\, $^{\text{\sf 3}}$Computer Engineering Department, Bilkent University, Ankara 06800, Turkey\, $^{\text{\sf 4}}$Computer Science and Engineering, Sabanci University, Istanbul 34956, Turkey}

\corresp{$^\ast$To whom correspondence should be addressed.}

\history{Received on XXXXX; revised on XXXXX; accepted on XXXXX}

\editor{Associate Editor: XXXXXXX}

\abstract{
\textbf{Motivation:}
Researchers need a rich trove of genomic datasets that they can leverage to gain a better understanding of the genetic basis of the human genome and identify associations between phenotypes and specific parts of DNA.
However, 
sharing genomic datasets that include sensitive genetic or medical information of individuals can {lead to} serious privacy-related consequences if data lands in the wrong hands. 
Restricting access to genomic datasets is one solution, but this greatly reduces their usefulness for research purposes. To allow sharing of genomic datasets while addressing these privacy concerns, several studies propose privacy-preserving mechanisms for data sharing. Differential privacy (DP) is one of such mechanisms that formalize rigorous mathematical foundations to provide privacy guarantees while sharing aggregated statistical information about a dataset. However, it has been shown that the original privacy guarantees of DP-based solutions degrade when there are dependent tuples in the dataset, which is a common scenario for genomic datasets (due to the existence of family members). \\
\textbf{Results:} In this work, we introduce a near-optimal mechanism to mitigate the vulnerabilities of the inference attacks on differentially private query results from genomic datasets including dependent tuples. 
We propose a utility-maximizing and privacy-preserving approach for sharing statistics by hiding selective SNPs of the family members as they participate in a genomic dataset. By evaluating our mechanism on a real-world genomic dataset, we empirically demonstrate that our proposed mechanism can achieve up to 40\% better privacy than state-of-the-art DP-based solutions, while near-optimally minimizing the utility loss.
\\
\textbf{Availability:} https://github.com/CMU-SAFARI/SNP-Selective-Hiding\\
\textbf{Contact:}\href{omutlu@ethz.ch}{ omutlu@ethz.ch},\href{otastan@sabanciuniv.edu}{ otastan@sabanciuniv.edu},\href{exa208@case.edu}{ exa208@case.edu}\\
\textbf{Supplementary information:} Supplementary data are available at \textit{Bioinformatics}
online.}

\maketitle

\section{Introduction} \label{IntroductionSec}


As technologies improve the cost and scale of sequencing, it has become possible to sequence genomes from large cohorts of patients. Today, researchers have access to large genomic datasets, whereby they can study associations between variants and complex traits.
However, as shown by earlier studies, the public availability of genomic data - even in anonymized form - raises serious privacy concerns~\citep{humbert2015anonymizing,gymrek2013identifying}. 
Hence, many institutions (i.e., data owners who collect genomic data), rather than publicly releasing their genomic datasets, provide limited access to these datasets through queries. Such queries typically seek to extract statistical information about the dataset (referred to as a "{\emph{statistical dataset}}"). They are formed and submitted by the researchers, computed at the data owner institution, and only the final results are shared with the querying researchers. 
One prominent example of such approach is the access to the results of {\emph{genome-wide association studies}} (GWAS) ~\citep{simmons2016enabling,jiang2014community,tramer2015differential, uhlerop2013privacy,yu2014scalable,johnson2013privacy}.  

Although this approach provides stronger privacy protection for the dataset participants, previous work has shown that such statistical genomic datasets are prone to {\emph{membership}} and {\emph{attribute inference attacks}}~\citep{backes2016membership}. An adversary, using the results of the queries, the genotype of a target, and {the} publicly available {\emph{minor allele frequencies}} (MAFs) of the {\emph{single nucleotide polymorphisms}} (SNPs) used in the study, can infer the membership of the target to the corresponding dataset (or to the case group of the corresponding GWAS)~\citep{homer2008resolving,wang2009learning}. This attack is considered serious because in most cases, dataset participants are associated with known sensitive information (e.g., cancer predisposition). 

{\emph{Differential privacy}} (DP)~\citep{dwork2008differential} is one of the privacy protection concepts that has received widespread popularity for sharing aggregate  statistics from human genomic datasets due to its theoretical guarantees~\citep{uhlerop2013privacy, yu2014scalable, johnson2013privacy}. Such that, even if there is {\emph{only}} one different tuple in two datasets (called {\emph{neighbouring datasets}}), it is 
hard to differentiate between the query results of these two datasets. The probability of distinguishing the results of the neighbouring datasets is controlled by a parameter called
{\emph{privacy budget}} $\epsilon$. However, DP has a known drawback as it makes no {assumption} about the {\emph{correlation}} between dataset tuples. This may degrade the privacy guarantees of DP and {give} the adversary a stronger ability to extract more sensitive information if the dataset includes dependent tuples, which is a common situation for genomic datasets as genomes of family members are correlated. 

Previous work {show} how the dependency between dataset tuples may reduce the privacy guarantees of DP~\citep{liu2016dependence,song2017pufferfish,zhao2017dependent} and {propose} general mechanisms to tackle this problem. Recently, \cite{almadhoun2019differential,almadhoun2020inference} analyze and show the privacy risk due to the inference attacks on differentially-private query results by exploiting the dependency between tuples in a genomic dataset. To mitigate this privacy risk, \cite{almadhoun2019differential} formalize the notion of $\epsilon$-DP for genomic datasets with dependent tuples to avoid the inference of sensitive information by any adversary with prior knowledge about the {tuples correlation}. 

However, to provide privacy guarantees for the dependent tuples in genomic datasets, existing DP-based solutions 
suggest changing the value of the privacy parameter $\epsilon$ (i.e., adding more noise to the released statistics depending on the number of dependent tuples and the strength of {relationship} between them). Such higher noise amounts significantly degrade the utility of the shared GWAS statistics, especially when the query results also include data from independent tuples in the dataset.
On the other hand, medical research necessitates highly accurate information for high quality and effective research outcomes. Therefore, it is also crucial to develop utility-preserving countermeasures for this privacy risk. 

In this work, we propose a novel privacy- and utility-preserving mechanism for sharing statistics from genomic datasets to attain privacy guarantees while taking into consideration the dependency between tuples. 
As discussed, the main reason for the aforementioned privacy risk is the existence of dependent tuples in the genomic datasets due to familial relationships. Therefore, our {\textbf{goal}} is to reduce the level of such dependency without significantly {weakening} the utility. 
To achieve this, inspired from our previous work~\citep{Kale2017AUM}, we propose an optimization-based countermeasure to {\emph{selectively}} {hide} genomic data of dataset participants to distort the dependencies (familial relations) among them without significantly degrading dataset responses, thus, the utility.

The {\textbf{key idea}} of our proposed {\emph{"selective hiding"}} mechanism 
is to {hide} some selected SNPs of family members (as they join to the genomic dataset) to (i) reduce the kinship relationship between them and (ii) keep the utility of {the} shared GWAS statistics high. 
By doing so, the constructed GWAS dataset includes only the obfuscated genomes of the dependent tuples. Thus, in case of a data breach, familial relationships between the GWAS participants are also protected. Also, the proposed method selectively hides {\emph{only}} the dependent tuples, keeping the genomes of independent tuples intact (which improves utility).

We assume that the GWAS dataset shares the kinship coefficients between its participants (e.g., as a part of its metadata) and a potential adversary uses this information along with the published GWAS statistics in order to infer sensitive attributes about the dataset participants. Even if metadata about the dataset is not shared, an adversary can infer the kinship coefficient between dataset participants by issuing several queries to the dataset. 
We evaluate the proposed algorithm against such an adversary by using real-life genomic datasets. 
Our results show that the proposed approach can provide both improved privacy and higher utility compared to existing work. 
As a result of the proposed countermeasure, dataset owners will share data {realizing} that the privacy of their participants will be protected. Also, individuals will be more open to donating their data to medical datasets for research knowing their privacy is uncompromised. Finally, researchers will know that they receive high-utility information from medical datasets.

The rest of this paper is organized as follows. Section 2 presents related prior works on genomics privacy, differential privacy mechanisms under dependent tuples, and our contributions. Section 3 describes the necessary background on protecting kinship inference and differential privacy. Section 4 explores our privacy threat model, followed by Section 5 which explains our approach. In Section 6  we evaluate our proposed strategy and compare it to the state-of-art mechanisms. Section 7 presents conclusions and highlights future research directions that are pointed by this paper.

\section{Related Work} \label{RelatedWorksSec}

In this section, we will summarize the state-of-the-art published studies on genomic privacy and differential privacy in particular. 

\subsection{Privacy of Genomic Data}

In recent years, privacy-preserving publishing of genomic data has received much
attention~\citep{erlich2014routes}. One of the widely-used promising privacy-preserving solutions is the {DP} framework. DP provides rigorous mathematical mechanisms for limiting the information leakage through adding noise to the statistics results in GWAS~\citep{simmons2016enabling,jiang2014community,tramer2015differential, uhlerop2013privacy,yu2014scalable,johnson2013privacy}.
Existing work basically utilizing the privacy guarantee of DP as a protective measure against inference attack scenario (e.g. membership attack discovered by~\citep{homer2008resolving}) even if the attacker has access to external auxiliary information.
\citep{yu2014scalable,johnson2013privacy,uhlerop2013privacy} proposed differentially-private algorithms to release the aggregate human genomic statistical results from genomic datasets as GWAS. Using a controlled amount of noise from \emph{Laplace distribution}~\citep{nissim2007smooth}, help enhancing privacy of all participants in a GWAS. In these algorithms, researchers submit genomic queries e.g cell counts, MAF, and ${\chi}^2$ statistics, and receive the query results in a privacy-preserving manner through DP algorithms. However, these proposed DP mechanisms assume that all the dataset tuples are independent,  which may degrade the privacy guarantees when such correlations exist between the tuples in the dataset. 

\subsection{Differential Privacy under Dependent Tuples}

The adversary can exploit auxiliary channels to get  information about the tuples correlation within the genomic dataset. \cite{kifer2011no} were the first to show this DP vulnerability. Therefore, they propose the Pufferfish framework \citep{kifer2012rigorous} as a generalization of DP to handle this threat.
Following the Pufferfish, different studies \citep{he2014blowfish,chen2014correlated, yang2015bayesian, cao2017quantifying, song2017pufferfish} provide perturbation mechanisms to handle the correlation between tuples for various applications. 
Recently, \cite{liu2016dependence} show that an adversary can utilize the pairwise dependencies within a location dataset to predict the participant's location from the differentially private query results~\citep{liu2016dependence}. To mitigate this privacy threat, \cite{liu2016dependence} propose a {\emph{Laplace}} mechanism defined as {\emph{dependent differential privacy}} (DDP) to tackle the pairwise {correlation} between any two tuples in the dataset.
To improve the privacy and utility guarantees of \citep{liu2016dependence}, \cite{zhao2017dependent}  present a new definition of {the} DDP, which can handle numeric and non-numeric queries, to address any adversary with arbitrary correlation knowledge.
Moreover, \cite{almadhoun2019differential, almadhoun2020inference} discuss attribute and membership inference attacks against differential privacy mechanisms, when the datasets include dependent tuples. As a countermeasure for these attacks, \cite{almadhoun2019differential} adjust the global sensitivity of the query before applying {\emph{Laplace perturbation mechanism}} (LPM) to the query results. 

\subsection{Contribution of This Work}

DP-based solutions to address the privacy risks due to the existence of dependent tuples in {statistical} datasets (including GWAS datasets) require addition of high noise values to the results of statistics queries, and hence they cause a significant loss in utility of the query responses. Here,  we propose a different approach to address the same problem. Our proposed solutions rely on selective masking of genomic {\emph{loci}} in a GWAS dataset to (i) decrease the estimated kinship coefficients between relatives in the dataset, (ii) provide privacy against an adversary that utilize correlations in the published statistics, and (iii) provide privacy for dataset participants (e.g., against kinship inference) in case the dataset {is} breached. Our results show that the proposed scheme provides both better privacy and higher utility than {the} existing solutions. 

\section{Background}

Here, we provide a brief background about our recent work on protecting kinship inference from public genomic datasets (which is the basis of the proposed algorithm) and differential privacy. 

\subsection{Protecting kinship inference from public genomic datasets}\label{sec:optimization}

In a previous work, we {define} two routes that leak kinship information from publicly available {datasets}~\citep{Kale2017AUM}. We {show} how {kin} relationship between participants of anonymous genomic {datasets} can be efficiently identified using (i) genotype similarity, and (ii) outlier allele pair counts. We {show} that the relatedness of two individuals can be inferred based on their genotype similarity using a kinship metric. We {observe} that such kinship metrics are mostly dominated by the number of SNPs that are {\emph{heterozygous}} in both individuals. Thus, before publicly sharing data, genomic positions wherein the two individuals are found to be heterozygous can be hidden as it decreases the kinship coefficient between two family members effectively. However, we also {show} that this alone will cause another privacy leakage as the number of positions where the two family members are heterozygous will be too small. Simply comparing this number to the population, one could infer that the two individuals are indeed in the same family. To mitigate these risks, in our earlier work ~\citep{Kale2017AUM} we {propose} a technique to protect kinship privacy against these risks while maximizing the utility of shared data. 
The method involves systematic identification of minimal portions of genomic data to mask as new participants are added to the {dataset}. Choosing the proper positions to hide is cast as an optimization problem in which the number of positions to mask is minimized subject to privacy constraints that ensure the familial relationships are not revealed. The privacy constraints are the privacy risks defined above. The former constraint pushes kinship values between family members to be equal to a preset value after the removal of SNPs. The latter one ensures that the number of heterozygous allele counts will not be too small hence, not become outliers in the {dataset} statistics.



\subsection{Differential privacy} \label{DP-Section}
Under the differential privacy, one's inclusion within a
dataset should make no statistical difference in an algorithm's
output. Therefore, two {datasets} that only differ by a {\emph{single}}
record should produce statistically similar results when
running a private algorithm. 
DP provides formal guarantees that applying a probabilistic mechanism $\mathbb{A}$ over two neighboring input datasets \textit{D} and \textit{{D}'}, which only differ by a single tuple, should make no big statistical difference in the distribution of query results $\mathbb{A}$(D) and $\mathbb{A}$(D)'. The degree of this difference can be controlled by privacy budget $\epsilon$. More formally, DP can defined as follows:

\begin{definition}{$\epsilon$-Differential Privacy~\citep{dwork2008differential}}

A randomized algorithm $\mathbb{A}$ achieves $\epsilon$-differential privacy if for any pair of neighboring datasets \textit{D}  and \textit{{D}'}, and any \textit{O $ \subseteq Range \left ( \mathbb{A} \right )$}, 
\end{definition}

\[\textit{Pr[$\mathbb{A}$(D) $\in$ O] $\leq$ $e^{\epsilon}$Pr[$\mathbb{A}$(${D}'$) $\in$ O]}\]

LPM is one famous instance dealing with numerical data to achieve differential privacy guarantees. It is based on adding noise from a Laplace distribution proportional to the query's global sensitivity (where the Laplace scale is equal to $\Delta Q$/$\epsilon$). Global sensitivity ($\Delta Q$) is the maximum possible change in the query outputs between datasets \textit{D}  and \textit{{D}'}~\citep{nissim2007smooth}.


\section{System and Threat Models} \label{ThreatModelSec}

The dataset owner maintains a statistical dataset ${D}$, and responds to users' statistical queries. 
To provide statistical information about the dataset in a privacy-preserved way, the dataset owner computes randomized query results $\mathbb{A}(D)$ using {LPM-based DP} (as in Section~\ref{DP-Section}), and sends it back to the users. The adversary in our scenario can be one of the users. The adversary can send various statistical queries to the dataset. In a recent work, we discuss the vulnerability of dependent tuples in a statistical dataset due to different statistical queries~\citep{almadhoun2020inference}. Here, for simplicity, we focus on a "count query", in which the adversary forms its query asking about the sum of values of a specific SNP $j$ among the dataset participants sharing the same demographic data, such as location or age (we assume an SNP value of $0$, $1$, or $2$, representing the number of its minor alleles). Limiting the scope of the query to a limited number of dataset participants allows the adversary to have a higher inference power about the sensitive genomic information {of} a target, especially if the query result is computed over the target and target's family members.  

This is a realistic attack scenario since statistical relationships between dataset participants are typically shared in the metadata of genomic datasets. According to~\cite{bennett2008standardized}, pedigree structures is a piece of metadata that is included in many genetic studies. 
Furthermore, there are several online public databases (e.g., Ysearch.org and SMGF.org) that collectively contain hundreds of thousands of surname-haplotype records, aiming at helping the public to identify their distant patrilineal relatives and the potential surnames of their biological fathers. 
With the availability of this information, considering an attribute inference attack (in which the goal of the adversary is to infer genomic data of a target dataset participant using the query results), we have the following assumptions for the adversary: 
\begin{itemize}
\item The adversary knows the membership information of all individuals in the dataset. 
\item The adversary knows the dependencies (e.g., kinship coefficient) between the individuals in the dataset. As discussed, the adversary can obtain this information from the metadata of the dataset. Alternatively, the adversary can also estimate the kinship coefficients between the dataset participants using the responses to its queries. 
\end{itemize}

\section{Proposed Work}

\begin{figure}[h]
\centering
\vspace{-5pt}
\includegraphics[width=0.48\textwidth]{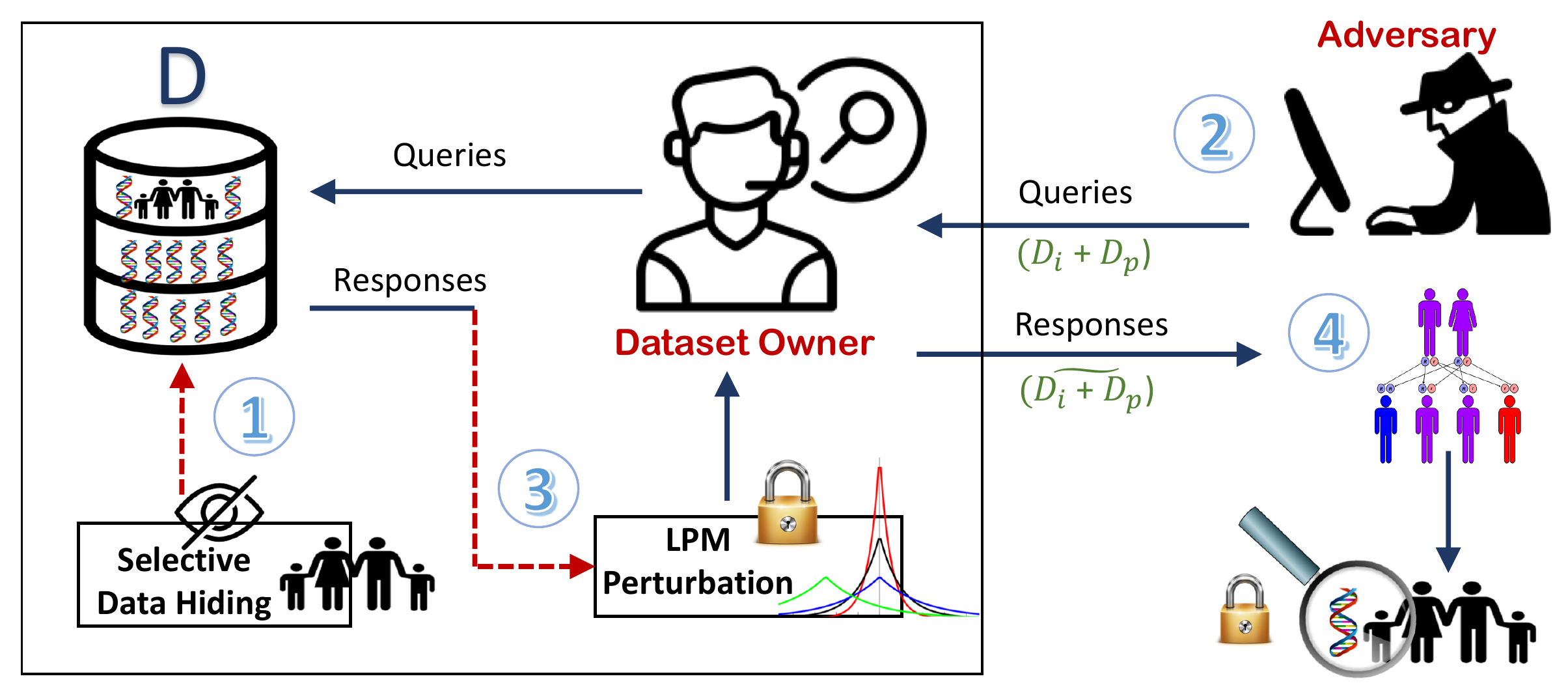}
\vspace{-15pt}
\caption{Our proposed model. 1) The dataset owner selectively hides SNPs from the family members included in the dataset during data collection. 2) The adversary sends the count queries to the dataset owner. 3) The dataset owner applies LPM to the query results and sends them to the adversary. 4) The adversary runs the attribute inference attack against the target $i$ by using (i) results of differentially-private count queries, (ii) dependency between the target and target's family members that are in the dataset $D$, and (iii) Mendel's law.}
\label{fig:01}
\end{figure}

Let dataset $\textit{D}$ includes ${\mathbb{N}}$ individuals and $\textit{m}$ SNPs. We assume a statistical query to the dataset is computed over $\textit{q}$ dataset participants, including a target $i$ and other  {$p$} dataset participants ($\textit{q = 1+p}$). 
$\textit{$D_{i}^{j}$}$ represents the value of SNP $j$ for target individual $i$ and $\textit{$D_{p}^{j}$}$ represents the sum of the SNP $j$ values for other $\textit{$(p)$}$ participants that are involved in the query computation. 
We let ($\delta$) be the added Laplace noise with scale 2/$\epsilon$.  
Set $\mathbf{F}$ ($\mathbf{|F|}= f$) includes individuals from the same family (i.e., target $i$ and his/her family members), and set $\mathbf{U}$ ($\mathbf{|U|}= u$) includes the other unrelated members (non-relatives) in the dataset. Note that there may be more than one family in the dataset and the privacy risk for each family can be shown similarly. Therefore, for the sake of simplicity, we assume the dataset includes {\emph{only}} one family. 
We show the overview of the proposed algorithm in Figure~\ref{fig:01}.

Similarly to the previous work~\citep{Kale2017AUM}, we assume family members share their data in a sequential  order. For each new incoming family member to the dataset, we hide some selected SNPs to decrease kinship coefficients among family members and preserve their familial privacy. The main differences of this work are: 
\begin{itemize}
\item The original selective sharing scheme in~\citep{Kale2017AUM} considers a publicly available dataset and it aims to reduce the kinship coefficients between the dataset participants to hide the familial relationships. Here, the statistical dataset is not public. Therefore, our aim is {\emph{not}} to specifically to hide  relation of the participants. Instead, our goal is to reduce the kinship coefficients so that (i) privacy vulnerability due to sharing of statistics computed over dependent tuples is minimized; and (ii) utility of the shared statistics still remain high. As a result, we also exclude the outlier constraints part (in Section~\ref{sec:optimization} and more details in ~\citep{Kale2017AUM} )  in the optimization model and focus on satisfying the kinship constraints only.  For completeness, below we describe the part of the formulation and the approach that we proposed in~\citep{Kale2017AUM} that is also used here.

\item We design the proposed method to hide {\emph{overlapping}} regions among the family members first, and solve the optimization later. The goal is to have better privacy and higher utility. 
\end{itemize}

To reduce the kinship coeffcient, we will hide positions based on their SNP configurations. Therefore, we use a notation to denote the positions with different SNP configurations for an individual  and a family. For an individual, $j$, a particular genomic position can have a SNP configuration $s_j$ where $s_j$ can take values in \{0,1,2\}. We denote the total number of positions the individual has with the SNP configuration of $s_j$ as $n_{s_j}$, i.e, $n_0$ is the number  of positions with SNPs value of 1, etc. This shows how many genomic locations are recessive homozygous, heterozygous, and dominant homozygous. When there is more than one person, we will refer to the number of genomic positions with a particular SNP configuration of the family members. For example, for a family of three, $n_{121}$ indicates the number of positions for which the first individual's SNP value is 1, the second's is 2, and the third's is 1. If $j$ is denoted with $*$ for any person, the person can have any of the SNP values.  

To calculate the kinship coefficient between two individuals $i$ and $k$, we use the robust  kinship estimator proposed by \cite{Manichaikul2010}:
\vspace{-5pt}
\begin{equation} \label{eq1}
\phi_{ik} =(2n_{11} - 4(n_{02} + n_{20}) - n_{*1} + n_{1*} ) / { 4n_{1*}} 
\end{equation}
when $n_{1*}$ < $n_{*1}$,  $k_{th}$ individual has more heterozygous positions than the $i_{th}$ member. $n_{11}$ presents the number of genomic position where both individuals are heterozygous. $n_{20}$ and  $n_{02}$ indicate the number of SNPs where the first individual (i) is homozygous dominant and the second individual (k) is homozygous recessive. 

Our solutions will find which positions to hide and this is decided based on the SNP configuration. We define a variable,  $x_{j}$, to denote the number of particular SNP configuration to be hidden from the latest arrived family member. 
Using the Equation \ref{eq1}, one can easily calculate $x_{11}$; the number of genomic positions to be removed in order to decrease the kinship coefficient down to a preset $\phi^\prime$ value between two individuals:

\begin{equation} \label{eq2}
x_{11} \!=\! \frac{2n_{11} \!-\! 4(n_{02}\!+\!n_{20}) \!-\! n_{1*} \!+\! n_{*1}(1 \!-\! 4 \phi_{ik}^ \prime))}{2(1\!-\! 2\phi_{ik}^\prime)} \enspace
\end{equation}

To have kinship coefficient lower than a preset $\Phi$, the problem can be cast as an integer programming problem as follows:

\vspace{-10pt}
\begin{multline}
\label{eq3}
\text{min} \quad x_{11} \\
\text{subject to } \\
 2n_{11} \!-\! 4(n_{02}\!+\!n_{20}) \!-\! n_{1**} \!+\! (1\!-\!4\Phi) n_{*1} \!\le \!(2\!-\!4\Phi) x_{11} \\
 x_{11}  \!\le\! n_{11}
\end{multline}
\vspace{-10pt}

\vspace{-10pt}
\begin{multline} 
\label{eq4}
\text{min} \quad x_{101} + x_{111} + x_{121} + x_{110} + x_{112} \\
\text{s.t. } \quad \\
 2n_{11*} \!-\! 4(n_{02*}\!+\!n_{20*}) \!-\! n_{1**} \!+\! (1\!-\!4\Phi) n_{*1*} \!\le \!(2\!-\!4\Phi) x_{11*} \!-\! x_{101}\! -\! x_{121} \\
 2n_{1*1} \!-\! 4 (n_{2*0} \!+\! n_{0*2}) \!-\! n_{1**} \!+\! (1\!-\!4\Phi) n_{**1} \!\le \!(2\!-\!4\Phi) x_{1*1} \!-\! x_{110}\!-\!x_{112} \\
 2n_{*11} \!-\! 4(n_{*02} \!+\! n_{*20}) \!-\! n_{**1} \! +\! (1\!-\! 4 \Phi) n_{*1*}\! \le \! (1\!-\!4\Phi) x_{11*} \!+\! 2x_{111} \!-\! x_{1*1} \\
 x_{11*} = x_{111} + x_{110} + x_{112} \\
 x_{1*1} = x_{111} + x_{101} + x_{121}\\
 x_{101}, x_{111}, x_{121}, x_{110}, x_{112} \in Z_{\ge 0}.
\end{multline}

\vspace{-5pt}
 The objective function in the integer programming model (\ref{eq4}) minimizes the number of SNP positions to be hidden subject to kinship constraints that are derived using kinship formula in \cite{Manichaikul2010}.   For families having more than two members, the optimization model considers all the pairwise kinship coeffcients among the related members. Equation \ref{eq4} shows the optimization model for a three member family.
 We use CPLEX (IBM Inc.) to solve the integer programming \citep{cplex2009v12}. 
  

The optimization model is solved every time when a new family member arrives at the dataset. First, we consider the overlapping SNP positions among the family members at the dataset. Once the number of positions and their configurations are determined by the optimization procedure, we select these positions from  their overlapped region., If the number of SNPs to hide is larger than the number of SNPs in the overlap, there is not enough SNP existing in the overlapping region, we run the model to remove the rest of SNPs from the latest arrived member.  As the dataset is not public, we assume that the dataset owner has information about the previously removed SNPs. Alternatively, after data collection is complete, the dataset owner can first identify the families, and then process the genomes one by one to identify the hidden part before the dataset is available for the statistical queries. 

\begin{figure}[H]
\includegraphics[scale=0.129]{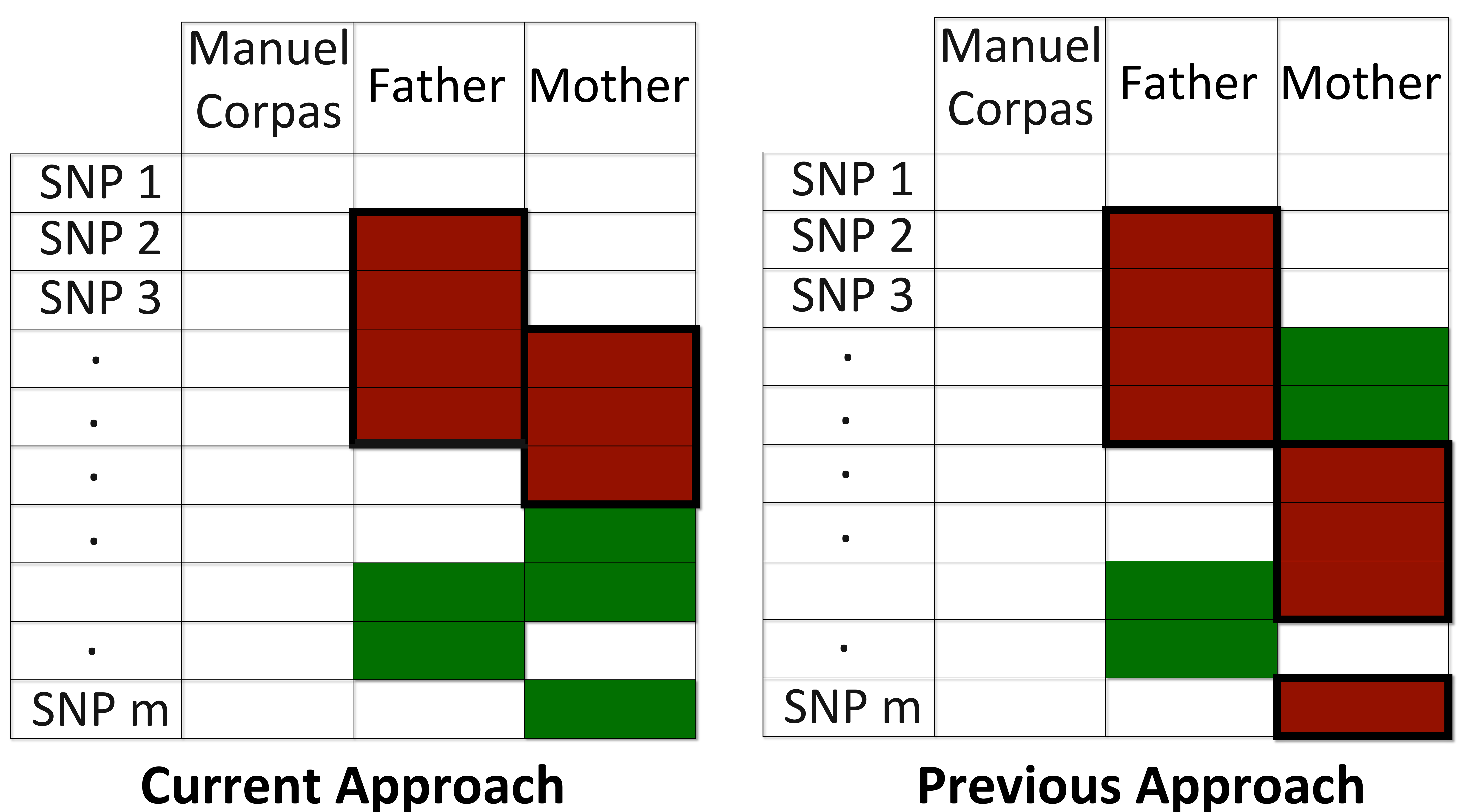}
\caption{Comparison of our proposed approach and the one in~\citep{Kale2017AUM}. The green-colored areas denote the available SNP positions to be hidden. Red-colored areas are the removed regions. In the proposed approach, we aim to hide from the region with maximal overlap.}
\label{fig:1}
\end{figure}

Hiding from the overlapping SNPs among the family members allows to (i) reduce the kinship estimates between multiple family members by hiding less number of SNPs, and hence have higher utility for the obfuscated dataset; and (ii) have higher privacy by having multiple hidden SNPs for a SNP position, and hence further confusing a potential adversary about the query results.  Figure~\ref{fig:1} shows how the new SNP set to be hidden is selected from the overlapping regions of previously hidden set.  Note that the adversary (who sends statistical queries to the dataset) cannot observe the hidden SNPs as the dataset is not published.

\begin{figure}[H]
\centering
\vspace{-5pt}
\includegraphics[width=0.3\textwidth]{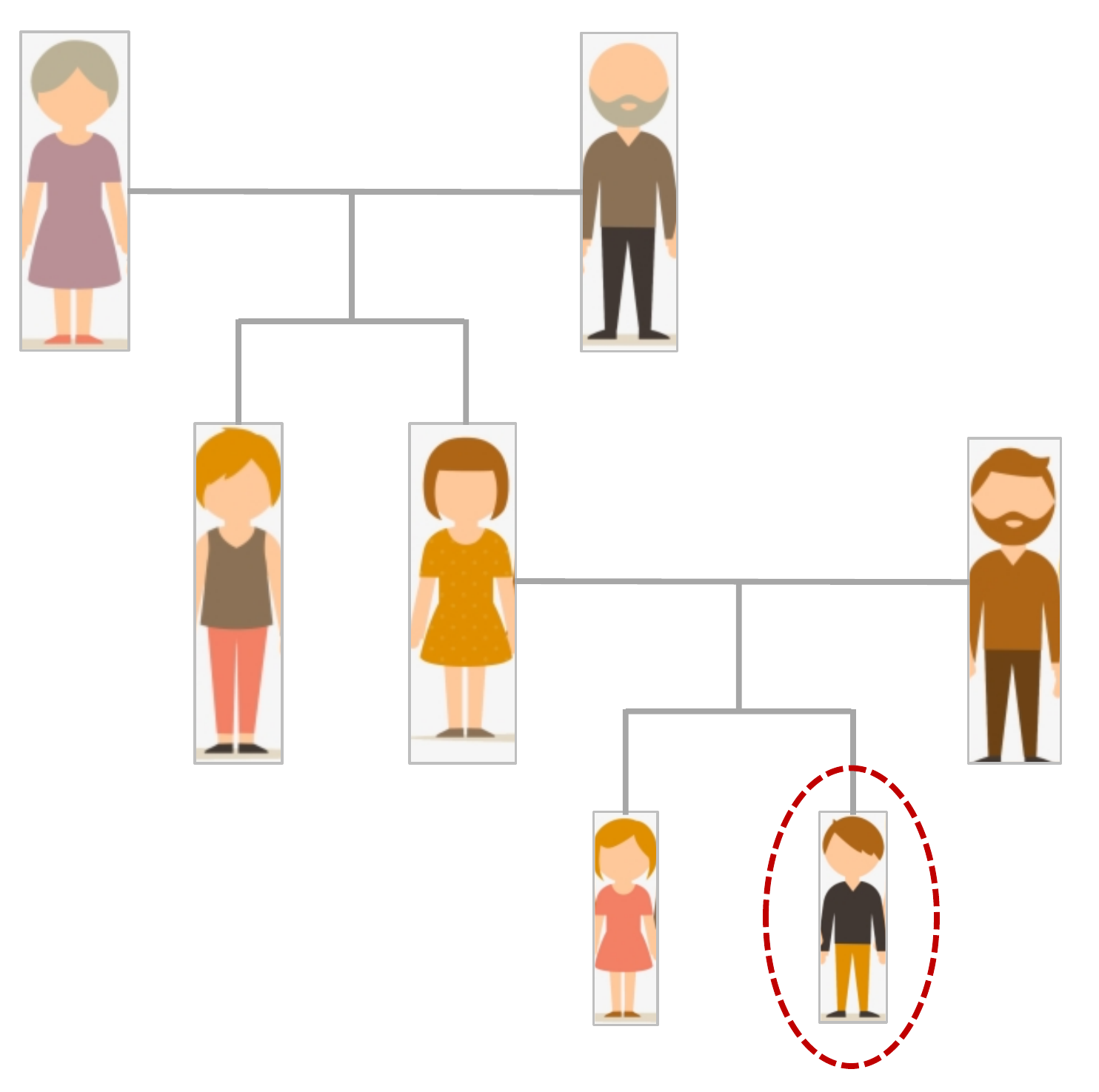}
\vspace{-15pt}
\caption{Manuel Corpas family tree.}
\label{fig:2}
\vspace{-15pt}
\end{figure}

In the following, we provide a toy example describing how the proposed selective hiding process work for the individuals in the {\emph{Manual Corpas}} family tree and we know the SNP data of the family and all the overlapping regions among them. \citep{corpas2013crowdsourcing} (Figure~\ref{fig:2}).
\begin{enumerate}

\item Manual Corpas arrives to the {dataset} (or his genome is processed the first). No SNPs are hidden from his genome.  
\item When the father arrives (or father's genome is processed), we first calculate the number of required SNPs to be hidden from the father using the optimization model with the aim of reducing the kinship between the son and the father. Then, we pick the required SNPs from the overlapping region, and the rest of SNPs are selected randomly.We hide these SNPs from the father. 
\item When the mother arrives, since we already removed the overlapping region before, her and the son's kinship coefficient is already decreased by one familial degree compared to their original value. No need to hide extra SNPs from the mother. (This steps shows the heuristic approach minimizes the random selection)
\item The aunt arrives. We run the optimization model for four people in such a way that kinship coefficients between both aunt-mother and aunt-son decrease, while preserving the decreased kinship coefficients in the previous steps. 
\end{enumerate}

After repeating this selective hiding process for each dataset participant, sequentially, all (required) records in the dataset becomes obfuscated and the dataset can now accept statistical queries. 
We consider the count query by the users (or the adversary). Following the attack scenario proposed by~ \cite{almadhoun2019differential}, to limit the number of dataset members included in the query results, the adversary sends its query specified by some demographic properties (e.g. age, address).
Dataset owner computes the result of the query on the dataset with missing SNPs (missing SNPs of some dataset participants are due to the proposed selective sharing algorithm).
Dataset owner reports (i) the query result (sum of all SNP values for the dataset participants that are considered in query computation) and (ii) number of dataset participants that are used to compute the query results ($\textit{q}$). Note that if a dataset participant is involved in the query computation,
but its corresponding SNP has been hidden (due to the proposed selective hiding algorithm), that participant still contributes to the number of dataset participants $\textit{q}$ that are used to compute the query result (i.e., from the adversary's point of view, the query is still computed over $\textit{q}$ individuals). 
In a response to a count query for a SNP $j$, the dataset owner computes a noisy query result $\textit{$\widetilde{D_{pi}^{j}}$}$, by adding Laplace noise with parameter 2/$\epsilon$.
The query result includes the sum of the SNP $j$ values for a target $i$ ($\textit{$D_{i}^{j}$}$) and other $p$ participants included in the query results ($\textit{$D_{p}^{j}$}$). We assume that the adversary has access to auxiliary information about the membership of each participant including the target $i$, and also to the familial relationship $\mathbb{R}$ between the target and other individuals in the dataset (that is computed over the obfuscated dataset with the hidden SNPs and released as metadata by the dataset owner). Hence, the adversary can infer the value of $\textit{$D_{i}^{j}$}$ for target $i$ using the {SNP values of} dependent people related to the target that is used to compute the query result, as shown in~\citep{almadhoun2020inference}. 

\section{Evaluation} \label{EvaluationSec}

To evaluate the privacy and utility performance of our proposed selective hiding algorithm, we {use} the correctness metric over a real-world genomic dataset to show the robustness of our mechanism. We next discuss our evaluation in details.

\subsection{Dataset Description} \label{DatasetDescriptionSec}

For the evaluation, our dataset $\textit{D}$ contains partial DNA sequences from two sources:
\begin{itemize}
   \item {1000 Genomes} phase 3 data~\citep{10002015global}
   \item Manuel Corpas Family Pedigree~\citep{corpas2013crowdsourcing} 
\end{itemize}

\subsubsection{1000 Genomes phase 3 data}
We {use} data from 1000 Genomes phase 3~\citep{10002015global}, to obtain data for the unrelated individuals from the same or different population of the target and his family members. We extracted the genotypes from chromosome 22 for 176 participants from the European population using the Beagle genetic analysis package~\citep{browning2018one} to convert the values of genotypes to 0, 1, or 2 according to the number of minor alleles for each SNP. 


\subsubsection{Manuel Corpas (MC) Family Pedigree}
Manuel Corpas~\citep{corpas2013crowdsourcing} released his and his family members' genomes for research purposes. The dataset contains the DNA sequences in variant call format (VCF) for the father, mother, son (Manuel Corpas), daughter, and aunt. The family tree of the individuals in this dataset is illustrated in Figure~\ref{fig:2}.
We {choose} the son to be the target and we used the genomic records of his first and second-degree family members (father, mother, and aunt).

We extracted the common SNPs from all MC family members and 1000 Genomes members for the evaluation of the proposed algorithm. Finally, we combined the family genomic data with the unrelated individuals.

\subsection{Evaluation Settings}

To evaluate the proposed countermeasure against the attribute inference attack, we defined a {\emph{case-control dataset}} $\textit{D}$. $\textit{D}$ includes $\mathbb{N}$ individuals ($\mathbb{N}$= 180) from European population from the 1000 Genomes project dataset and MC family, in which (${\mathbb{N}\over 2}= 90$) are cases and (${\mathbb{N}\over 2}= 90$) are controls. 
As discussed in {Section}~\ref{ThreatModelSec}, the adversary aims to infer $m$ SNPs for a target $i$ using the results of queries over dataset $D$. Here, we assume that the adversary knows the kinship coefficients of the dataset participants (e.g., from the metadata of the dataset). Note that kinship coefficients shared by the dataset are computed after the proposed selective sharing algorithm (reflecting the actual kinship coefficients in the final dataset), and hence they are obfuscated to provide robustness.

\subsection{Evaluation Metrics} \label{EvaluationMetricsSec}
 
To evaluate the performance of the proposed algorithm against attribute inference attack, we used the correctness metric. Using the notion of the expected {\emph{estimation error}}, the {\emph{correctness}} of the adversary quantifies the distance {($\textit{Dist}$)} between the true value of {the} SNP and {the inferred} value of the SNP for the target individual between (i) $D_{i}^{j}$, which is the true value of SNP $\textit{j}$ for the target individual $\textit{i}$ and (ii) {$\tilde{D_{i}^{j}}$}, which is the inferred value of SNP $\textit{j}$ for the target individual $\textit{i}$ by the adversary. We compute the correctness for all $m$ targeted SNPs of the target $i$ as follows:
\begin{equation} \label{equ:error}
\mathbb{C} = 1 - \sum_{j=1}^{m}P\left (D_{i}^{j} \mid \tilde{D_{pi}^{j}} \right )\left | Dist\left (D_{i}^{j}, \tilde{D_{i}^{j}} \right ) \right |,
\end{equation}

To quantify the {\emph{utility loss}} due to the proposed mechanism, we calculate the average change in the actual query result $D_{pi}^{j}$ and the noisy query result $\tilde{D_{pi}^{j}}$ for all $m$ targeted SNPs as follows:
\begin{equation} \label{equ:utilty}
\mathbb{U} = \frac{1}{m} \sum_{i=1}^{m} |Dist(\frac{D_{pi}^{j}}{2q}, \frac{\tilde{D_{pi}^{j}}}{2q})|,
\end{equation}


\subsection{Experimental Results} \label{ExperimentalResultsSec}
In an inference attack, we assume the differentially private query results are computed by accounting for: (i) target $i$ and multiple first and second-degree family members in $\mathbf{F}$; and (ii) target $i$, multiple family members in $\mathbf{F}$, and multiple other unrelated members (non-relatives) in $\mathbf{U}$. 
We evaluate the performance of the attack under two assumptions:
\begin{itemize}
	\item Independent assumption (w/o dep): the adversary assumes that there is no correlation between the participants in $D$.
	\item Dependent assumption (w/ dep): the adversary utilizes the familial relationships between the participants in $D$ to perform the genome reconstruction for target $i$.
\end{itemize} 

We also compare the proposed algorithm with the one proposed in~\citep{almadhoun2019differential}, which aims to adjust the {\emph{privacy parameter}} of DP to provide privacy guarantees for the dependent tuples in the dataset. 
According to~\citep{almadhoun2019differential,almadhoun2020inference}, if all the tuples in the dataset are independent, then the noisy query output achieves DP with the same privacy budget $\epsilon$. 
However, if the dataset includes dependent tuples, one needs to augment the scale of Laplace noise using a smaller $\epsilon$ value (or a larger query sensitivity) to achieve DP. 
Using the notion of a leaked information ratio for different privacy budgets $\epsilon$,~\citep{almadhoun2019differential} adjust the global sensitivity of the query to mitigate the information leaks resulting from the attribute inference attack. 

In the following, we (i) compare the dependent and independent assumptions to show the vulnerability due to independent assumption, (ii) show the performance of our proposed mitigation algorithm (by hiding selective SNPs from the family members) against an adversary that uses the dependencies in its attack, (iii) hide random SNPs (without using any optimization) from the family members rather than selective hiding, to show the benefit of selective hiding, and (iv) compare the proposed mitigation algorithm with the one in~\citep{almadhoun2019differential} to asses the proposed algorithm. 
\begin{figure}[h]
\centering
\vspace{-5pt} 
\includegraphics[width=0.495\textwidth]{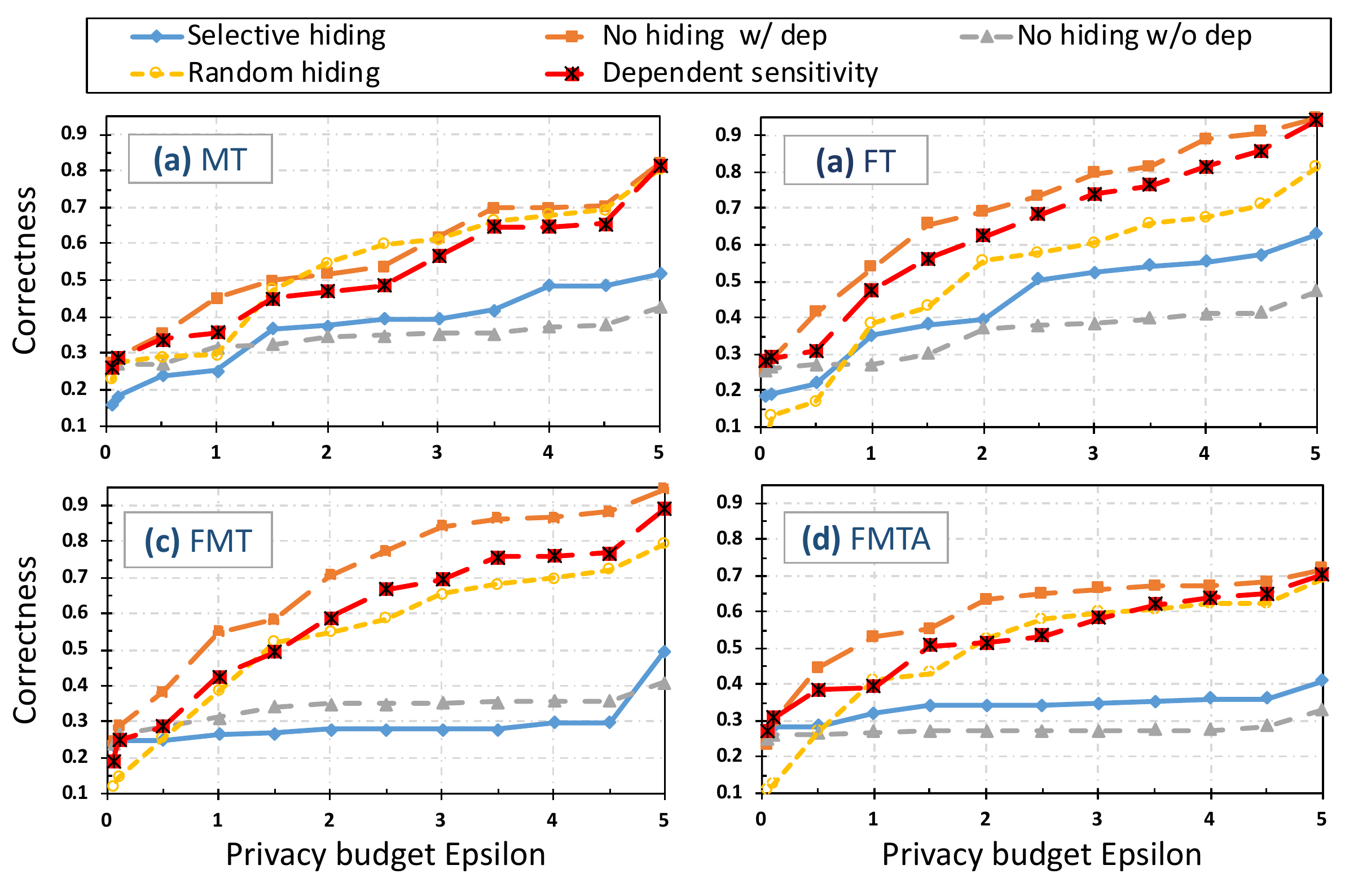}
\vspace{-15pt}
\caption{The effect of different values of the privacy budget, $\epsilon$, on the adversary's correctness in inferring the targeted SNPs, considering different number of family members in $\mathbf{F}$ ($|\mathbf{F}|=f$) included in the noisy results of count query. The query results include (a) MT: mother and target, (b) FT: father and target, (c) FMT: father, mother and target (d) FMTA: father, mother, target, and aunt.}
\label{fig:02}
\end{figure} 
 \begin{figure*}[ht]
\centering
\includegraphics[width=0.8\textwidth]{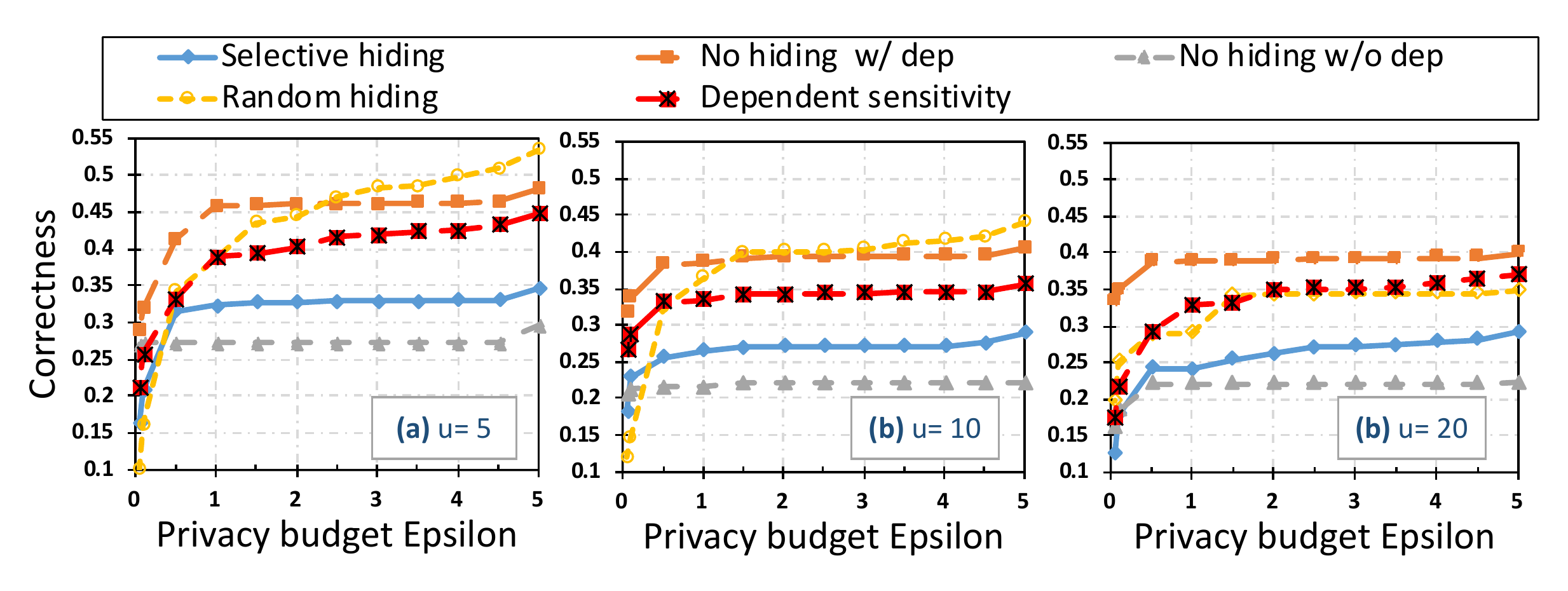}
\vspace{-15pt}
\caption{The effect of different values of the privacy budget, $\epsilon$, on the adversary's correctness in inferring the targeted SNPs, considering 2 first-degree relatives (father and mother) with different numbers of non-relatives in $\mathbf{U}$ ($|\mathbf{U}|=u$) included in the noisy results of count query. The query results include 5, 10, and 20 unrelated members in (a),(b), and (c) respectively.}
\label{fig:03}
\vspace{-10pt}
\end{figure*}

\subsubsection{Privacy Performance} \label{PrivacyPerformanceSec}

In Figure~\ref{fig:02}, we evaluate the effect of different values of the privacy budget, $\epsilon$, on the adversary's correctness in inferring the targeted $m$ SNPs. We also analyze the robustness of our proposed mechanism to the inference attack and compare it with the most similar existing work~\citep{almadhoun2019differential}. Here the query results include the statistics from the family members only. We start including 1 first-degree family member with the target $i$. First, we include the mother to the query results as in Figure~\ref{fig:02}(a), then we include the father of the target  as in Figure~\ref{fig:02}(b)). Third, we include both the father and the mother in the query results, as in Figure~\ref{fig:02}(c). Last, we consider a second-degree family member (aunt of target $j$) in the query results along with the father and the mother of the target {(Figure}~\ref{fig:02}(a)). 

Using the results of count queries over the case-control dataset $D$, we make the following key observations:
(i) The correctness of the adversary with the knowledge of the data dependency is up to $50\%$ more compared to the case in which the adversary does not consider the data dependency in the query results (Figure~\ref{fig:02}).
{(ii)}In accordance with the results of~\cite{almadhoun2020inference}, the most accurate inference of the adversary is achieved when the query computation includes target $j$ along with his father and mother (Figure~\ref{fig:02}(c)). Including a second-degree family member as in (Figure~\ref{fig:02}(d)) can enlarge the range of possible SNP values for the target, and hence make it more difficult to accurately infer the correct SNP value with a high probability.

(iii) Proposed selective hiding mechanism achieves better privacy for various privacy budgets, compared to the random hiding for different family members included in the query results, as illustrated in Figure~\ref{fig:02}.

Figure~\ref{fig:03} shows the effect of different values of the privacy budget, $\epsilon$, on adversary's success in terms of its correctness in inferring $m$ SNPs of target $i$.  We increase the number of non-relatives (from 5 to 20) that are included in the query computation along with first-degree family members of the victim. 
From these experimental results we make the following key observations:

(i) In accordance with our previous observations in Figure~\ref{fig:02}, the probability of inferring the true value of the targeted $m$ SNPs slightly increases (mostly 2\%-20\%) depending on the knowledge of the adversary about the dependency between tuples, as the value of the privacy budget, $\epsilon$, increases from 0.1 to 5. Hence, even when  including different number of non-relatives in the query results (e.g., the size of $\mathbf{U}$ changes from 5 to 20), there is a significant increase in the correctness of the adversary if the adversary has the knowledge of the data dependency, as shown Figure~\ref{fig:03}. However, in Figure~\ref{fig:03}, we observe that the difference between the correctness of the inferred SNPs with and without the knowledge of the data dependency is about 3 times less than when the query results include data for only family members of target $i$ (Figure~\ref{fig:02}). 

(ii) Applying our proposed countermeasure by selectively hiding the family members' SNP values is superior to the dependent sensitivity mechanism in terms of correctness metric.  Compared to the optimal DP privacy guarantees, in which we consider all the tuples to be independent {((No hiding w/o dep) in Figure}~\ref{fig:03}), our proposed mechanism achieves ($\sim$5\%) less privacy, while dependent sensitivity mechanism achieves ($\sim$15\%) less privacy guarantees under the same privacy budget, $\epsilon$. 

(iii) Randomly hiding the SNPs of the family members results in achieving less privacy guarantees, even if we compare it with the correctness results of the attribute inference attack, where no hiding method is applied (e.g. no hiding w/ dep in (Figure~\ref{fig:03}(a) and (b) for privacy budget, $\epsilon > 2.5$). 
\begin{figure}[h]
\vspace{-10pt}
\centering
\includegraphics[width=0.3\textwidth]{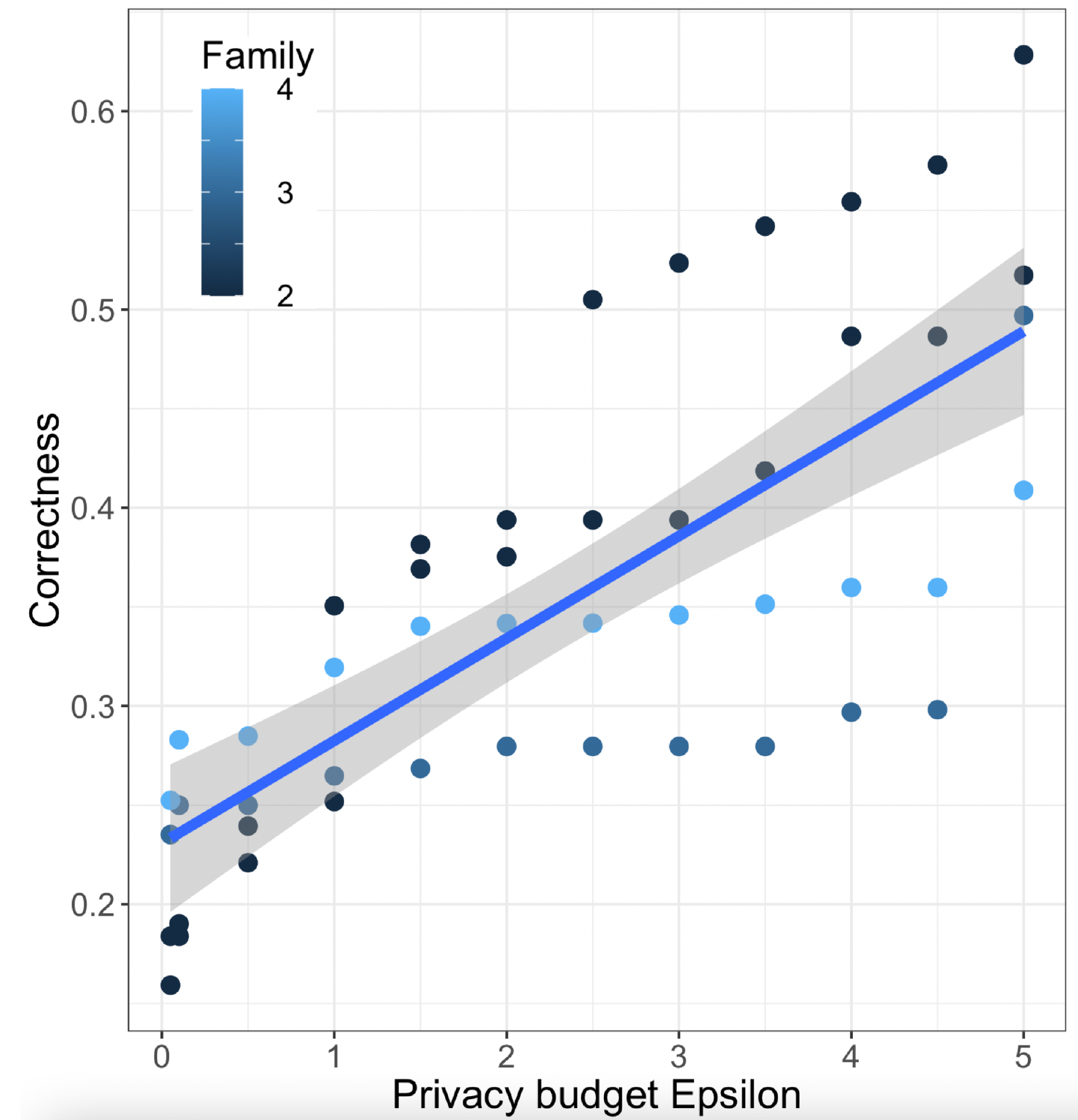}
\vspace{-10pt}
\caption{The effect of different values of the privacy budget, $\epsilon$, on the adversary's correctness in inferring the targeted SNPs, using different number of family members in $\mathbf{F}$ ($|\mathbf{F}|=f$) included in the noisy results of count query.}
\label{fig:04}
\vspace{-10pt}
\end{figure}

\begin{figure}[h]
\vspace{-10pt}
\centering
\includegraphics[width=0.495\textwidth]{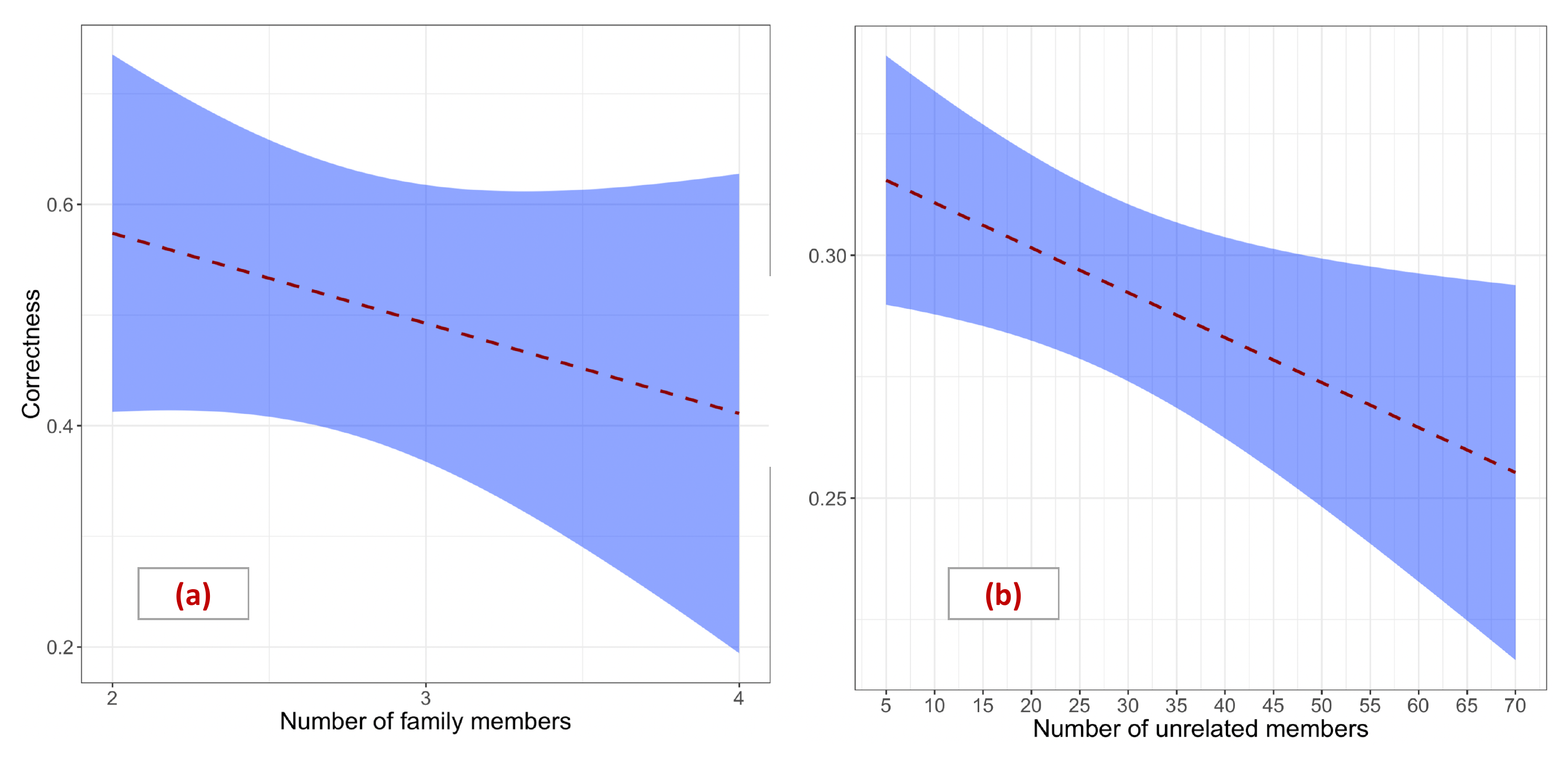}
\vspace{-15pt}
\caption{The relationship between different numbers of (a) family members in $\mathbf{F}$ ($|\mathbf{F}|=f$ and (b) non-relatives in $\mathbf{U}$ ($|\mathbf{U}|=u$) included in the noisy results of count query, and the adversary's correctness in inferring the targeted SNPs.}
\label{fig:05}
\vspace{-10pt}
\end{figure} 
Next, Figure~\ref{fig:04} shows the effect of different values of the privacy budget, $\epsilon$, used in DP, on the correctness of the adversary, when we apply selective hiding mechanism for family members SNPs, considering different number of family members to be included in the query results. The results illustrate the association between the privacy budget, $\epsilon$, and the correctness of the adversary for inferring the {\emph{actual}} values of the targeted $m$ SNPs. The probability of inferring the correct values increases significantly (by 30\%) as the budget privacy, $\epsilon$, increases from 0.1 to 5, as shown in Figure~\ref{fig:04}. This is expected as the more  $\epsilon$ values we use in the LPM-based DP, the less the added noise, and hence increasing the success of the inference attack.

Finally, we explore the robustness of the selective hiding mechanism for different number of related and unrelated people in the query results, without applying differential privacy. Figure~\ref{fig:05} show the relationship between the number of family members (as in Figure~\ref{fig:05}(a)) or the number of non-relatives (as in Figure~\ref{fig:05}(b)) in the query results and the probability of inferring the true SNPs value by the adversary, when we apply selective hiding mechanism. The results show that increasing the number of family members or unrelated individuals included in the query result, using selective hiding mechanism slightly decreases the correctness of the adversary, thus improving the privacy.

\subsubsection{Utility Performance} \label{UtilityPerformanceSec}
\begin{figure}[h]
\centering
\includegraphics[width=0.495\textwidth]{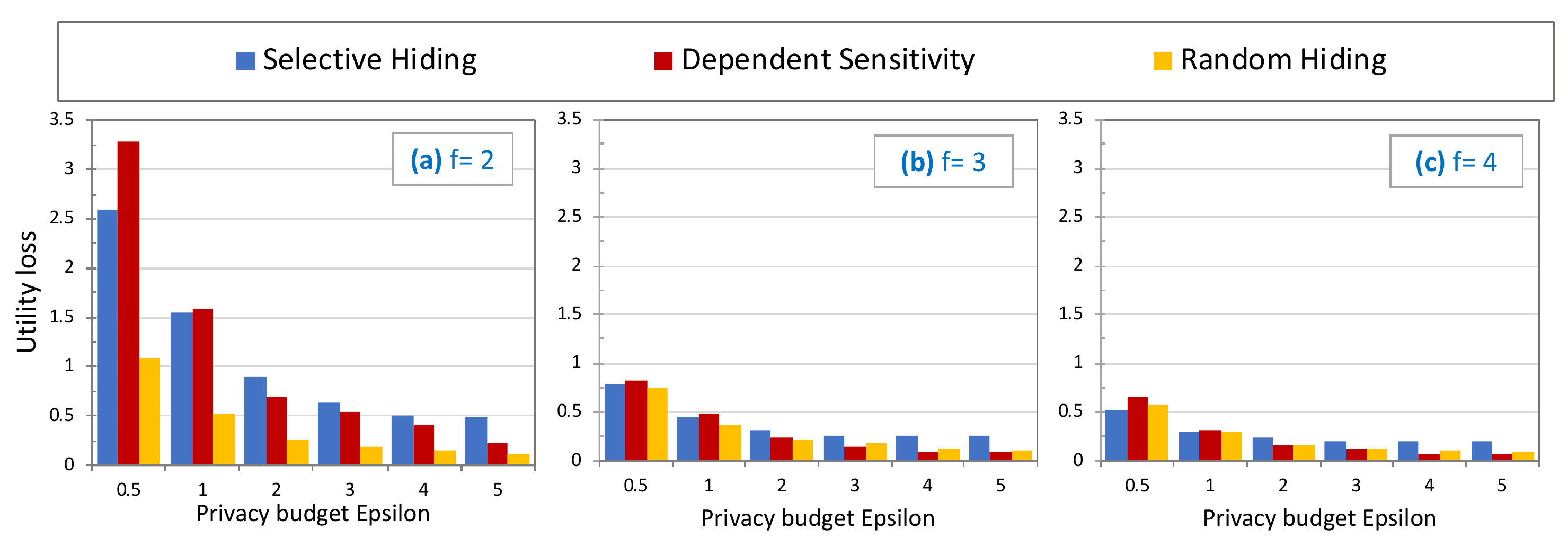}
\vspace{-15pt}
\caption{The effect of different values of the privacy budget, $\epsilon$, on the utility loss caused by applying different mechanisms as countermeasures against the attribute inference attack, using different number of family members in $\mathbf{F}$ ($|\mathbf{F}|=f$) included in the noisy results of MAF query.}
\label{fig:06}
\vspace{-10pt}
\end{figure}

\begin{figure}[h]
\centering
\includegraphics[width=0.495\textwidth]{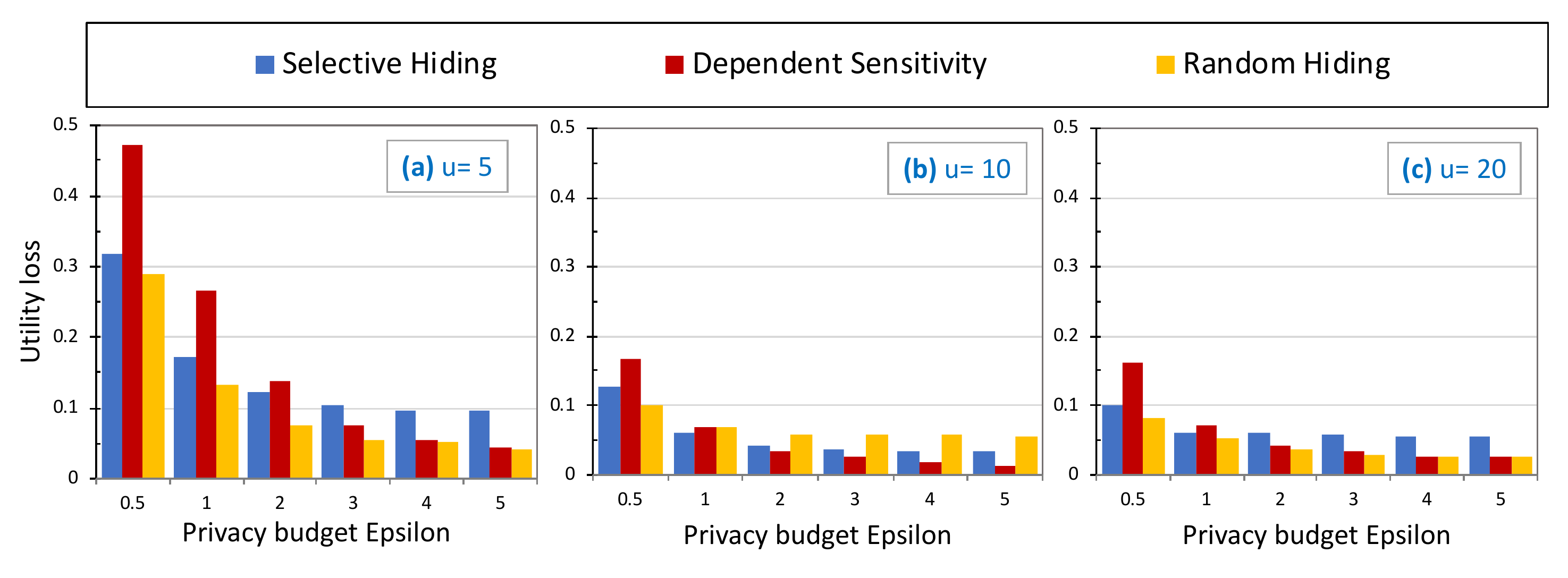}
\vspace{-15pt}
\caption{The effect of different values of the privacy budget, $\epsilon$, on the utility loss caused by applying different mechanisms as countermeasures against the attribute inference attack, using different number of non-relatives in $\mathbf{U}$ ($|\mathbf{U}|=u$) included in the noisy results of MAF query.}
\label{fig:07}
\vspace{-10pt}
\end{figure}

Publishing statistics of genomic datasets results in utility gain for the society as a whole. However, publishing these statistics could also result in privacy loss for the participants of the dataset, especially if the dataset includes correlated tuples. Hence, the goal of our proposed mechanism is to ensure that the privacy loss is restricted to an acceptable level, without causing a high loss in the potential utility gain, when compared with the case of publishing the original statistical results.
In the following, we compare our proposed mechanism with the existing dependent sensitivity countermeasure proposed in~\cite{almadhoun2019differential} and random hiding mechanism in terms of utility, using a MAF query over a dataset $D$ with m=100 SNPs.
Figure~\ref{fig:06} and Figure~\ref{fig:07} show the utility loss caused by hiding selective SNPs from the family members participating in the dataset $D$ and then adding noise to achieve ${\epsilon}$-DP by considering the dependence between tuples.

As in Section~\ref{PrivacyPerformanceSec}, we consider the query results to include the statistics from the family members only (Figure~\ref{fig:06}). Then, we calculate the utility performance of the three mechanisms considering query results with different numbers of unrelated individuals (Figure~\ref{fig:07}).  
Using the utility metric introduced in Section~\ref{EvaluationMetricsSec}, the results show that with smaller $\epsilon$ values, utility loss caused by the three mechanisms decreases.

The main idea of the dependent sensitivity mechanism~\citep{almadhoun2019differential} is augmenting the Laplace noise by decreasing the privacy budget, $\epsilon$, value to achieve DP for any dataset with dependent tuples. Our proposed mechanism adds a significantly smaller amount of noise, when $\epsilon \leq 1$, and hence provides better utility. For example, when ${\epsilon}=0.5$, and the query results include 5 unrelated individuals along with the family members (Figure~\ref{fig:07}(a)), the amount of utility loss caused by our mechanism is $33\%$ of utility loss caused by the dependent sensitivity.





\section{Conclusion} \label{ConclusionSec}

Developing new privacy-preserving techniques that facilitate sharing the outcomes of human genomic studies is necessary. The main goal of such techniques is to preserve the privacy of dataset donors without undermining the utility of the dataset, and hence the research outcomes. 
Differential privacy-based data perturbation techniques have known privacy limitations while sharing statistics from genomic dataset that contains dependent tuples. 
In this paper, we have proposed a ``selective hiding'' mechanism to mitigate the privacy risks caused by the correlations between the dataset tuples. 
We have evaluated our perturbation mechanism over a real-world genomic datasets and proved that it can achieve high privacy guarantees, while minimizing the utility loss. 
Our results show that the proposed scheme achieves both significantly better privacy and utility than the existing DP-based mechanisms.

\bibliography{bioinf-sample}

\end{document}